\documentclass[11pt, a4paper]{google}

\usepackage[authoryear, sort&compress, round]{natbib}
\usepackage{listings}
\usepackage{xcolor}
\usepackage{subcaption}
\usepackage{float}

\title{Building Browser Agents: Architecture, Security, and Practical Solutions}

\author{Aram Vardanyan\\
Founder, FillApp\\
\texttt{aram@fillapp.ai}}

\begin{abstract}
Browser agents enable autonomous web interaction but face critical reliability and security challenges in production. This paper presents findings from building and operating a production browser agent. The analysis examines where current approaches fail and what prevents safe autonomous operation. The fundamental insight: model capability does not limit agent performance; architectural decisions determine success or failure. Security analysis of real-world incidents reveals prompt injection attacks make general-purpose autonomous operation fundamentally unsafe. The paper argues against developing general browsing intelligence in favor of specialized tools with programmatic constraints, where safety boundaries are enforced through code instead of large language model (LLM) reasoning. Through hybrid context management combining accessibility tree snapshots with selective vision, comprehensive browser tooling matching human interaction capabilities, and intelligent prompt engineering, the agent achieved approximately 85\% success rate on the WebGames benchmark across 53 diverse challenges (compared to approximately 50\% reported for prior browser agents and 95.7\% human baseline).
\end{abstract}

\begin{document}

\maketitle

\section{Introduction}
\label{sec:introduction}

Browser agents powered by LLMs automate interaction with web interfaces. Several AI-native browsers and web-agent frameworks report strong performance on selected benchmarks \citep{xie2024osworld, koh2024visualwebarena} and show multi-step workflows such as shopping or travel planning. However, as Section~\ref{sec:background} discusses, these systems provide limited evidence about safe and reliable production operation.

A gap exists between demonstration and production operation. Findings from building and operating a production browser agent reveal that LLM capability is not the limiting factor in widespread adoption. Modern LLMs have sufficient reasoning ability to navigate web tasks effectively when provided with appropriate context and tools. The fundamental challenges are architectural: how the system manages context, what information the agent accesses, how safety boundaries are enforced, and whether to target generalization or specialization.

These challenges reflect a fundamental mismatch. The web was designed for humans, beings with vision, spatial awareness, selective attention, and the ability to process dynamic content including animations, videos, and layered interfaces. Current browser agents can replicate many human interactions such as clicks, typing, and scrolling. However, they struggle with aspects that are straightforward for humans: understanding visual hierarchy, processing time-dependent information, maintaining relevant context while discarding noise, and operating safely in environments where a single wrong action can cause irreversible damage.

This paper presents findings from production experience, not laboratory benchmarks. The insights come from observing where agents succeed, where they fail, and what architectural patterns separate reliable systems from demonstrations.

Among these challenges, security represents the most immediate barrier to autonomous operation. As Section~\ref{sec:prompt-injection} details, public security analyses of AI-native browsers and browser extensions demonstrate that prompt injection attacks remain effective even after multiple rounds of mitigation \citep{shapira2025mindtheweb, cohen2025fuzzing}. Studies consistently show non-trivial attack success rates in scenarios where agents have access to sensitive user data across multiple domains. When an agent operates with full user privileges across banking, email, and corporate systems, even a 1\% vulnerability rate represents unacceptable risk.

Reliable, safe browser agents require better architectural decisions instead of better LLMs. Production experience reveals four insights: context management determines success, architecture matters more than LLM scale, security requires programmatic constraints over LLM-based judgments, and specialization outperforms general-purpose approaches. Each insight challenges conventional assumptions about autonomous web agents.

The remainder of this paper examines each of these dimensions in detail, drawing from 
production experience to provide actionable guidance for building browser agents that are both 
capable and safe enough for real-world deployment.

\section{Background \& Related Work}
\label{sec:background}

\subsection{AI browsers and computer-use APIs}

A new class of AI browsers integrates large language models (LLMs) directly into the browsing experience. OpenAI's ChatGPT Atlas, launched in October 2025, embeds ChatGPT as a browser sidebar with an optional Agent Mode capable of performing autonomous tasks such as trip planning or shopping over the user's tabs \citep{openai2025atlas}. Perplexity Comet, announced in July 2025, functions as an AI browser that automates research, email triage, and other workflows from within the interface \citep{perplexity2025comet}.

Anthropic's computer use capability extends Claude models with tools to control a desktop environment and browser, separate from standalone AI browsers. The initial release in October 2024 exposed tools such as \texttt{click}, \texttt{type}, and \texttt{navigate}, with a focus on security mitigations and permission prompts \citep{anthropic2024computeruse}. Anthropic later introduced Claude for Chrome, a browser extension that brings this automation into existing sessions. Early reports highlighted productivity benefits and documented remaining prompt injection vulnerabilities \citep{anthropic2024chrome}.

Several open-source frameworks target this problem space. WebVoyager introduces a multimodal browser agent that combines page screenshots with text to complete tasks on real websites, achieving approximately 59\% task success on the WebVoyager benchmark \citep{he2024webvoyager}. The Browser Use framework builds on this work and reports 89.1\% success rate on the same benchmark across 586 tasks \citep{browseruse2024}. These systems show that LLM-driven browser agents can reach high success rates on static benchmarks, but provide limited visibility into production reliability, long-running workflows, and operational safety at scale.

\subsection{Browser automation MCPs and accessibility-based agents}

The Model Context Protocol (MCP) ecosystem now includes standardized servers for browser automation. Microsoft's Playwright MCP exposes a headless browser controlled via accessibility tree snapshots instead of raw screenshots. This approach provides the LLM with a structured view of roles, labels, and focusable elements, mapping these to executable actions \citep{playwrightmcp2024}. Chrome DevTools MCP provides access to the Chrome DevTools protocol, including performance traces and full accessibility trees, allowing agents to operate using semantic element references instead of coordinate-based interactions \citep{chromedevtoolsmcp2024}.

These MCP servers illustrate a design pattern that decouples the perception layer (accessibility tree, page snapshots, or screenshots) from the execution layer (tool calls such as \texttt{click}, \texttt{type}, \texttt{navigate}, \texttt{performance\_start\_trace}). Accessibility-driven representations simplify element selection and enable safety policy enforcement at the tool layer. The architecture presented in this paper follows this principle but extends it with hybrid vision and accessibility context, element reference versioning, bulk actions, and domain-specific safety constraints derived from production usage.

\subsection{Prompt injection and agent security}
\label{sec:prompt-injection}

Prompt injection represents the central security concern for tool-using agents. Systematic evaluations across multiple models and attack patterns have demonstrated high success rates, highlighting the need for defense-in-depth approaches \citep{liu2024prompt}.

Regarding AI browsers, Brave's security team demonstrated indirect prompt injection attacks against Perplexity Comet. Attackers hid adversarial instructions in elements invisible to the user, such as white text on white backgrounds or HTML comments. These instructions caused Comet to execute sensitive cross-site actions, including fetching one-time passwords from email or accessing banking portals, when the user asked it to "summarize this page" \citep{brave2025comet}. Later analyses by security vendors confirmed that such AI browsers bypass traditional browser security boundaries like the same-origin policy by executing cross-domain actions on behalf of the user.

Defenses that rely on detecting attacks, such as filtering suspicious inputs or attempting to separate user instructions from untrusted content, typically reduce the probability of attack success without eliminating attack classes. For agents with high-impact privileges such as email, banking, and corporate SaaS, even minimal failure rates remain unacceptable.

\subsection{Position of this work}

Most prior work on browser agents and AI browsers has focused on benchmark performance (WebVoyager, leaderboards) or security analysis in controlled environments. These studies decouple utility and safety from production constraints including latency budgets, cost, and complex user behavior.

This paper contributes an additional perspective by reporting on the year-long production operation of a browser agent used on real-life workflows, with full access to authenticated web sessions. The analysis focuses on architectural decisions that enable production usability: hybrid accessibility and vision context representations, execution-layer and tool design, memory management and intelligent context trimming strategies, and programmatic safety boundaries that enforce specialization. It combines production observations into patterns and constraints to inform the design of future AI browsers and secure agent architectures.

\section{The Human-AI Gap}
\label{sec:human-ai-gap}

The web was designed around human perception and motor control, not browser agents that act through structured representations and discrete actions. Humans combine vision, spatial awareness, selective attention, and continuous pointer control to navigate complex web pages. They track loading indicators, animated transitions, hover effects, and integrate these with working memory about the task goal. Sound signals state changes through notification tones, error chimes, and video audio.

Modern browser agents perceive pages through screenshots or structured representations of DOM elements and control the browser through discrete tool calls instead of continuous motion or direct audio interaction. They issue individual tool calls such as \texttt{click}, \texttt{type}, \texttt{scroll}, and \texttt{navigate}. This execution model enables agents to complete workflows including account sign-up, profile editing, and checkout or booking flows with few steps and limited conditional logic. However, these interactions approximate only a subset of what humans do naturally when interacting with a web page.

There is a deeper gap in sensing and timing. Humans interpret videos, animations, and micro-interactions such as progress bars or pulsing error states and adjust their timing based on interface responsiveness. They follow moving elements with the mouse and adapt to latency spikes or race conditions. Current browser agents typically lack integrated audio perception and can fail to align clicks with moving targets or rapidly changing layouts, leading to missed actions or incorrect states.

This gap is visible in interactive tasks that rely on real-time coordination across modalities. Drawing tools, fast-paced web games, or interfaces with rapid state changes are straightforward for humans but remain difficult for systems that operate only on occasional, static views of the page and discrete tool calls. These environments highlight how much human perception and timing still exceed current browser agents.

These differences drive specific architectural choices for production browser agents. Systems that lean only on vision-based screenshots struggle with dense interfaces and overlapping layers. Systems that rely only on structured views of page elements and relationships miss visual cues that humans use constantly, including motion, relative distance, and off-screen context. In this work, page representation and execution layer capabilities serve as central mechanisms for narrowing these human--AI gaps while staying within practical limits on latency and cost.

\section{Context Management}
\label{sec:context-management}

Context management determines whether browser agents remain reliable under realistic latency and token budgets. A browser agent must access enough information to understand the current page and plan actions, yet must avoid overwhelming the LLM with repetitive or irrelevant details. Poor context choices increase latency and token consumption. They also make failures more likely when the LLM loses track of relevant elements in a long conversation history.

Two design dimensions determine how browser agents manage context: how the system perceives the page and how it maintains conversation history over time. On the perception side, systems commonly choose between vision-based representations, text-based structured document representations, or hybrids that combine both. On the temporal side, the system either accumulates all intermediate snapshots and tool outputs as raw messages or compresses them aggressively into a lightweight history. Observations from production operation and evaluation on WebGames (Section~\ref{sec:production-validation}) support hybrid page representations paired with compressed history as a configuration that scales to long-running workflows in this architecture.

\subsection{Vision-based approaches}

Vision-based approaches present the agent with page screenshots and rely on the LLM's vision capabilities to identify targets and decide where to click. This representation resembles how human users perceive a web page: a visible viewport with clear spatial relationships, explicit layering, and immediate feedback on hover states, dialogs, and overlays. For simple layouts with a modest number of interactive elements, modern multimodal LLMs can reliably identify buttons, links, and form fields in a screenshot and issue corresponding click or type actions.

However, vision-only interaction exposes several limitations. Not all models support precise bounding box prediction for small or densely packed elements. Even models with built-in bounding box interfaces frequently misidentify small targets such as calendar cells or icon grids in date pickers. Gemini 2.5 Pro and Gemini 3 Pro expose vision APIs that return bounding boxes \citep{google2025geminivision}, yet still fail to place boxes reliably on dense grids of small elements. These errors accumulate on interfaces with many similar targets, where a single missed bounding box leads to repeated attempts and wasted actions.

\begin{figure}[H]
	\centering
	\begin{subfigure}{0.3\linewidth}
		\centering
		\includegraphics[width=\linewidth, alt={Date picker interface with small 24px cells}]{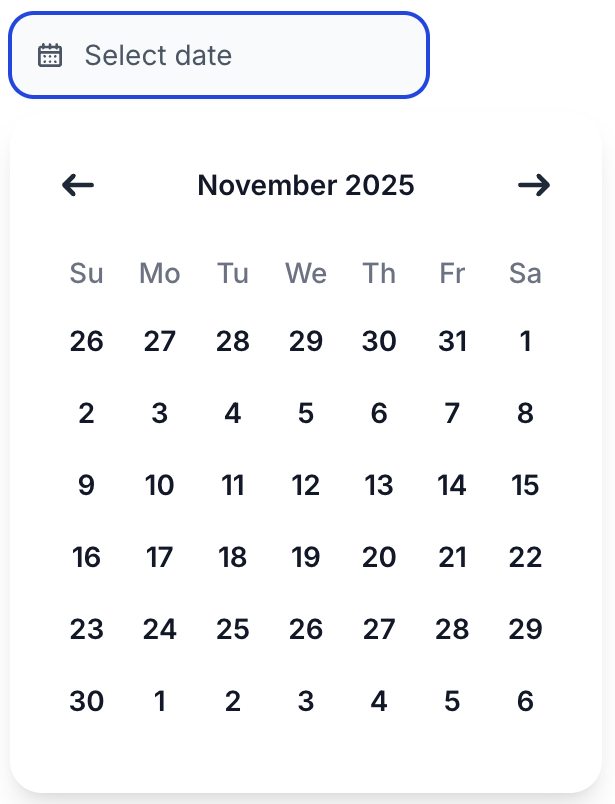}
	\end{subfigure}%
	\hspace{0.02\linewidth}
	\begin{subfigure}{0.26\linewidth}
		\centering
		\includegraphics[width=\linewidth, alt={Example of a dense date picker interface}]{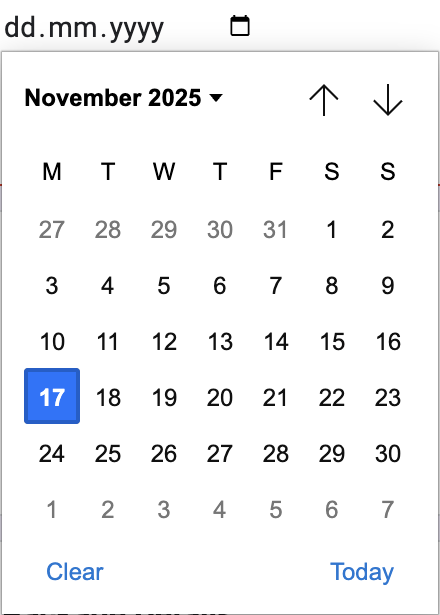}
	\end{subfigure}
	\caption{\textbf{Date picker examples with dense clickable elements.} Typical date picker interfaces where each date cell is only 24px, creating challenges for vision-based approaches to accurately identify and click specific dates.}
	\label{fig:date-picker}
\end{figure}

These limitations are visible in commercial AI browsers. Atlas, for example, does not disclose its internal implementation, but testing in October 2025 showed behavior consistent with a screenshot-driven vision agent. On date picker widgets similar to those in Google Forms (Figure~\ref{fig:date-picker}), the agent frequently failed to select the correct day, sometimes requiring several minutes of repeated attempts on a simple calendar interaction. The challenge in these cases is that vision models must place a bounding box precisely over a specific 24px cell in a tightly packed grid, a task that current vision models handle unreliably.

Another problem with screenshot-based interaction is that it works only with HTML elements for which the browser maintains bounding boxes. Parts of the web such as Google Sheets, Figma, and Canva operate primarily on HTML5 canvas, which does not expose individual interactive elements to the DOM or accessibility APIs.

To reduce reliance on model-driven bounding boxes, some systems overlay annotation layers on top of screenshots. The execution layer first parses the DOM, computes bounding boxes for clickable elements, and renders those boxes on the screenshot with numeric labels. The LLM then refers to element references such as ``click 5'', and the execution layer maps the chosen number back to the underlying DOM node.

\begin{figure}[H]
	\centering
	\includegraphics[width=1\linewidth, alt={Dashboard screenshot with numeric annotations on buttons}]{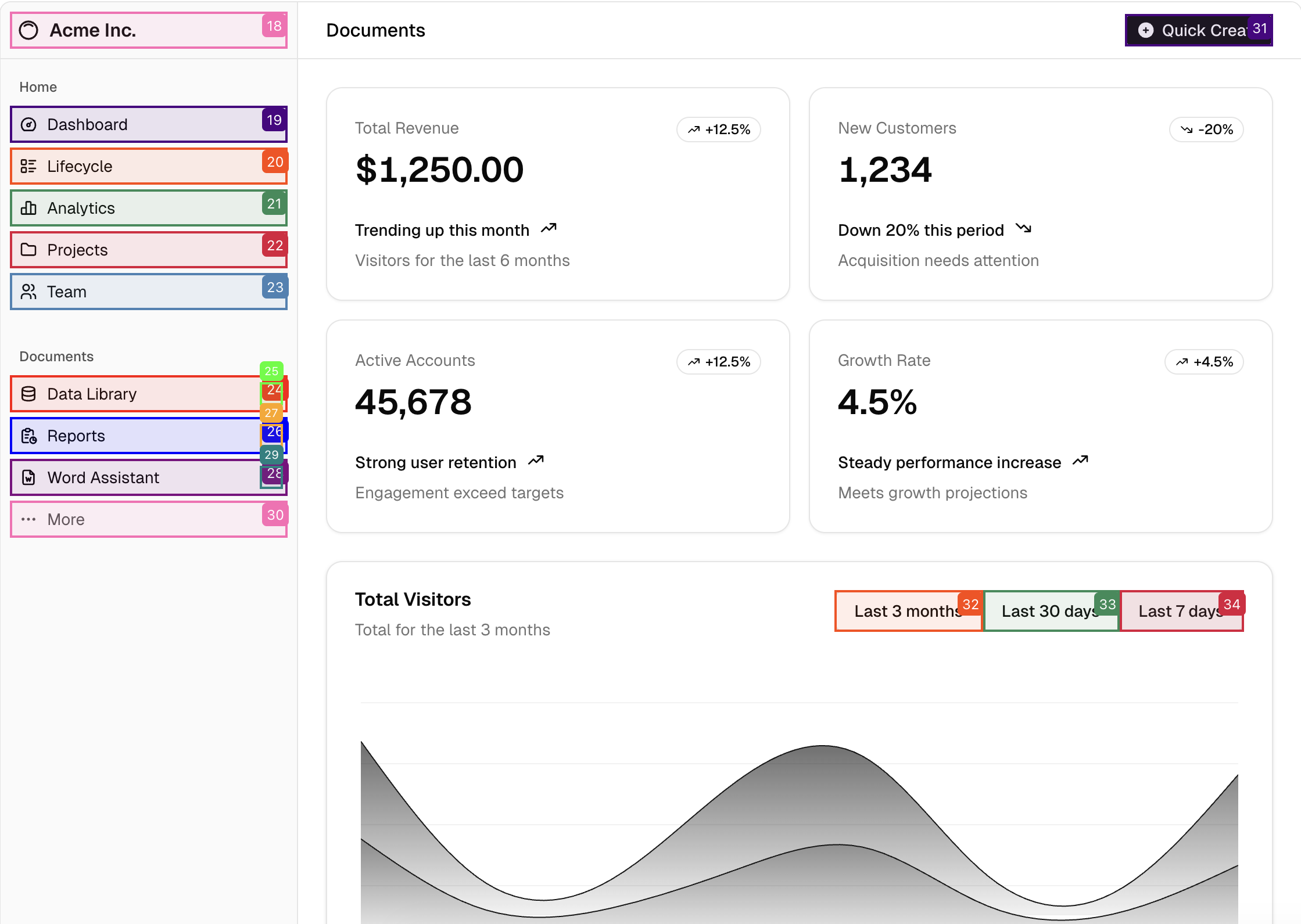}
	\caption{\textbf{Annotated screenshot approach for simple interfaces.} Clickable elements are labeled with numbers, allowing the LLM to reference them directly (for example, ``click button 5''). This works well when there are relatively few clickable elements.}
	\label{fig:dashboard-annotations}
\end{figure}

This annotation-based approach works well when there are only a handful of interactive elements. However, dense layouts make annotation unreliable. When a page contains hundreds of clickable elements, such as a date picker where each 24px cell is clickable, the overlay becomes cluttered and labels overlap, making the screenshot effectively unreadable to the LLM.

\begin{figure}[H]
	\centering
	\includegraphics[width=1\linewidth]{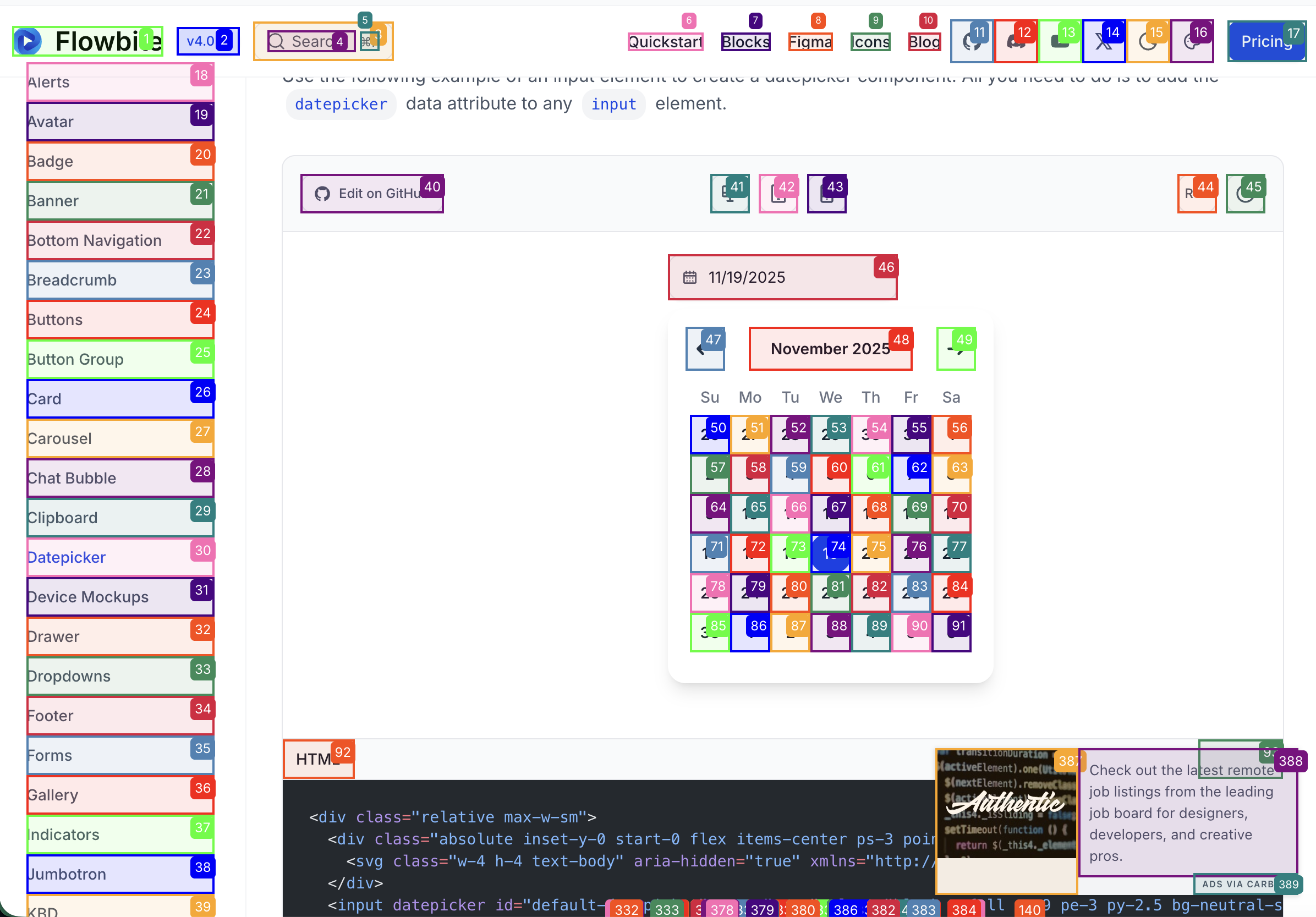}
	\caption{\textbf{Annotation approach fails with dense interfaces.} When attempting to annotate a date picker with many small clickable elements (24px each), the overlay becomes cluttered and confusing, making it difficult for the LLM to understand and reference specific elements.}
	\label{fig:messy-datepicker}
\end{figure}

\subsection{Text-based structured representations}

An alternative to vision is to send the LLM a text-based representation of the page structure. The most straightforward approach is to extract the page's DOM and pass it directly to the LLM. For simple pages with minimal markup, this can work: the LLM receives a tree of HTML elements, identifies interactive controls, and issues tool calls referencing those elements.

However, raw DOM quickly becomes impractical for realistic web applications. Modern pages contain deeply nested containers, styling hooks, decorative elements, and verbose ARIA attributes that obscure the semantic structure. For complex forms, it may take more than 100 lines of HTML to describe a single logical field, with multiple wrapper divs and accessibility annotations scattered across the markup. A simple text input labeled ``Total Weight (kg)'' can require over 100 lines of nested HTML elements (Figure~\ref{fig:html-markup}), making it difficult for the LLM to identify the label, control, and validation message.

\begin{figure}[H]
\centering
\begin{minipage}{\linewidth}
\begin{lstlisting}[language=HTML]
<div jscontroller="sWGJ4b" jsaction="EEvAHc:yfX9oc;" class="geS5n Jj6Lae">
  <div class="z12JJ">
    <div class="M4DNQ">
      <div
        id="i23"
        class="HoXoMd D1wxyf RjsPE"
        role="heading"
        aria-level="3"
        aria-describedby="i27"
      >
        <span class="M7eMe">Total Weight (kg)</span>
        <span class="vnumgf" id="i27" aria-label="Required question"> *</span>
      </div>
      <div class="gubaDc OIC90c RjsPE" id="i24"></div>
    </div>
  </div>
  <div jscontroller="oCiKKc" class="AgroKb">
    <div class="rFrNMe k3kHxc RdH0ib yqQS1 zKHdkd" jscontroller="pxq3x">
      <div class="aCsJod oJeWuf">
        <div class="aXBtI Wic03c">
          <div class="Xb9hP">
            <input
              type="text"
              class="whsOnd zHQkBf"
              autocomplete="off"
              tabindex="0"
              aria-labelledby="i23 i26"
              aria-describedby="i24 i25"
              required=""
              dir="auto"
            />
            <div class="ndJi5d snByac" aria-hidden="true">Your answer</div>
          </div>
          <!-- ... multiple wrapper divs ... -->
        </div>
      </div>
    </div>
  </div>
  <div id="i25" role="alert">
    <div class="rRld8e">
      <!-- ... decorative icon elements ... -->
    </div>
    <span class="RHiWt">This is a required question</span>
  </div>
</div>
\end{lstlisting}
\end{minipage}
\caption{\textbf{HTML markup for a simple form field.} A simple text input labeled ``Total Weight (kg)'' requires over 100 lines of complex HTML with nested divs, ARIA attributes, and decorative elements, far too verbose to pass efficiently to an LLM.}
\label{fig:html-markup}
\end{figure}

A more effective approach relies on the browser's built-in accessibility APIs. Browsers maintain an accessibility tree that represents the page in a form designed for screen readers and other assistive technologies. This tree strips away decorative markup, resolves ARIA relationships, and exposes only the semantic structure: roles, labels, descriptions, focus state, and interactive elements. Modern MCP-based tools such as Playwright MCP and Chrome DevTools MCP leverage these APIs to generate accessibility tree snapshots \citep{chrome2024devtools}. Instead of forwarding verbose HTML, they request structured nodes from the accessibility tree and encode each node with its role, label, description and focus state.

\begin{figure}[H]
	\centering
	\includegraphics[width=1\linewidth, alt={Chrome Accessibility Tree view of the form field}]{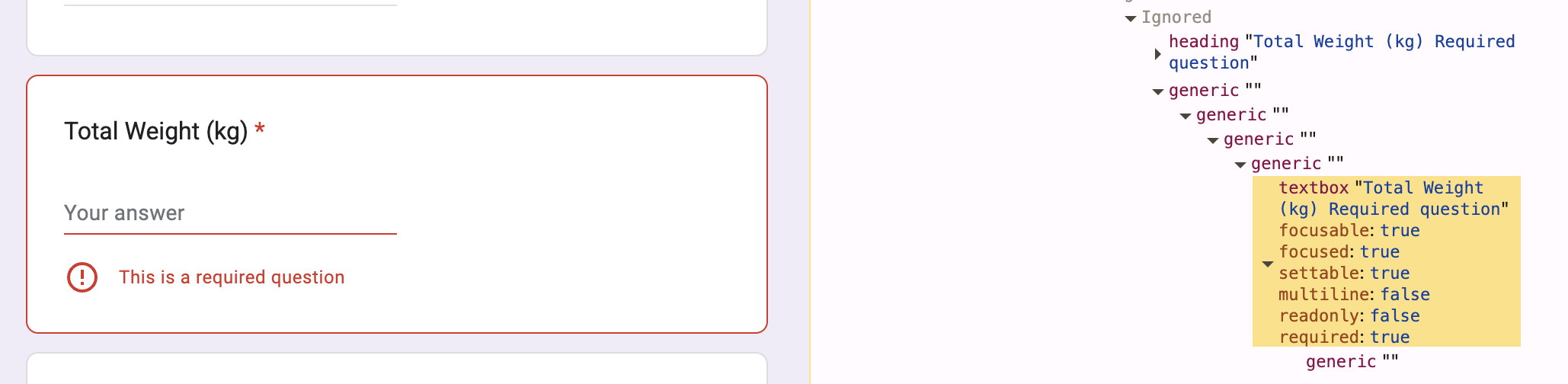}
	\caption{\textbf{Chrome's Accessibility Tree representation.} The browser's accessibility API provides a structured view of the page, exposing semantic information about elements without the verbose HTML markup.}
	\label{fig:chrome-accessibility}
\end{figure}

The same ``Total Weight (kg)'' field that required over 100 lines of HTML can be represented in just a few lines of accessibility tree snapshot text, as shown below. This compact representation omits structural noise while preserving all the information needed for the agent to understand and interact with the field. Figure~\ref{fig:chrome-accessibility} illustrates how the browser's accessibility tree provides this structured view. To type into the field, the agent issues a structured tool call referencing the element by its element reference:

\begin{figure}[H]
\centering
\begin{minipage}{\linewidth}
\begin{lstlisting}
ref=38 heading "Total Weight (kg) Required question" level="3" description="Required question"
ref=39 textbox "Total Weight (kg) Required question" description="This is a required question" focusable focused required
\end{lstlisting}
\end{minipage}
\end{figure}

\begin{figure}[H]
\centering
\begin{minipage}{\linewidth}
\begin{lstlisting}[language=Java]
type({
  ref: 39,
  text: "Hello World!"
})
\end{lstlisting}
\end{minipage}
\end{figure}

The accessibility tree snapshot makes element references explicit and stable enough for short interaction sequences. However, accessibility tree snapshots are not a universal solution. Many production web properties are not fully compliant with the Web Content Accessibility Guidelines (WCAG) \citep{w3c2018wcag, webaim2024million}, not because developers intentionally omit accessibility features, but because they did not invest the additional effort required to make interfaces fully accessible. Even well-known form builders and content management systems expose unlabeled buttons that control dynamic components such as date pickers or dropdowns. Others render tooltips and helper text at the end of the \texttt{<body>} element without semantic references linking them back to the target element, a common pattern used to work around CSS layering challenges. In these cases, the accessibility tree snapshot either hides important interactions or presents them in an order that is difficult for the LLM to interpret.

Layering further complicates snapshot-only interaction. Dialogs, overlays, and nested modals often stack multiple clickable elements on top of one another. If the snapshot does not reflect visual z-order or occlusion correctly, the agent may click background elements while a dialog is open or fail to identify which control will open a hidden dropdown or date picker. These gaps lead to designs that do not choose between screenshots and accessibility tree snapshots but rather combine them.

\subsection{Limitations of Grid-Based Mapping}

\begin{figure}[H]
	\centering
	\includegraphics[width=1\linewidth, alt={Dashboard screenshot overlaid with a coordinate grid}]{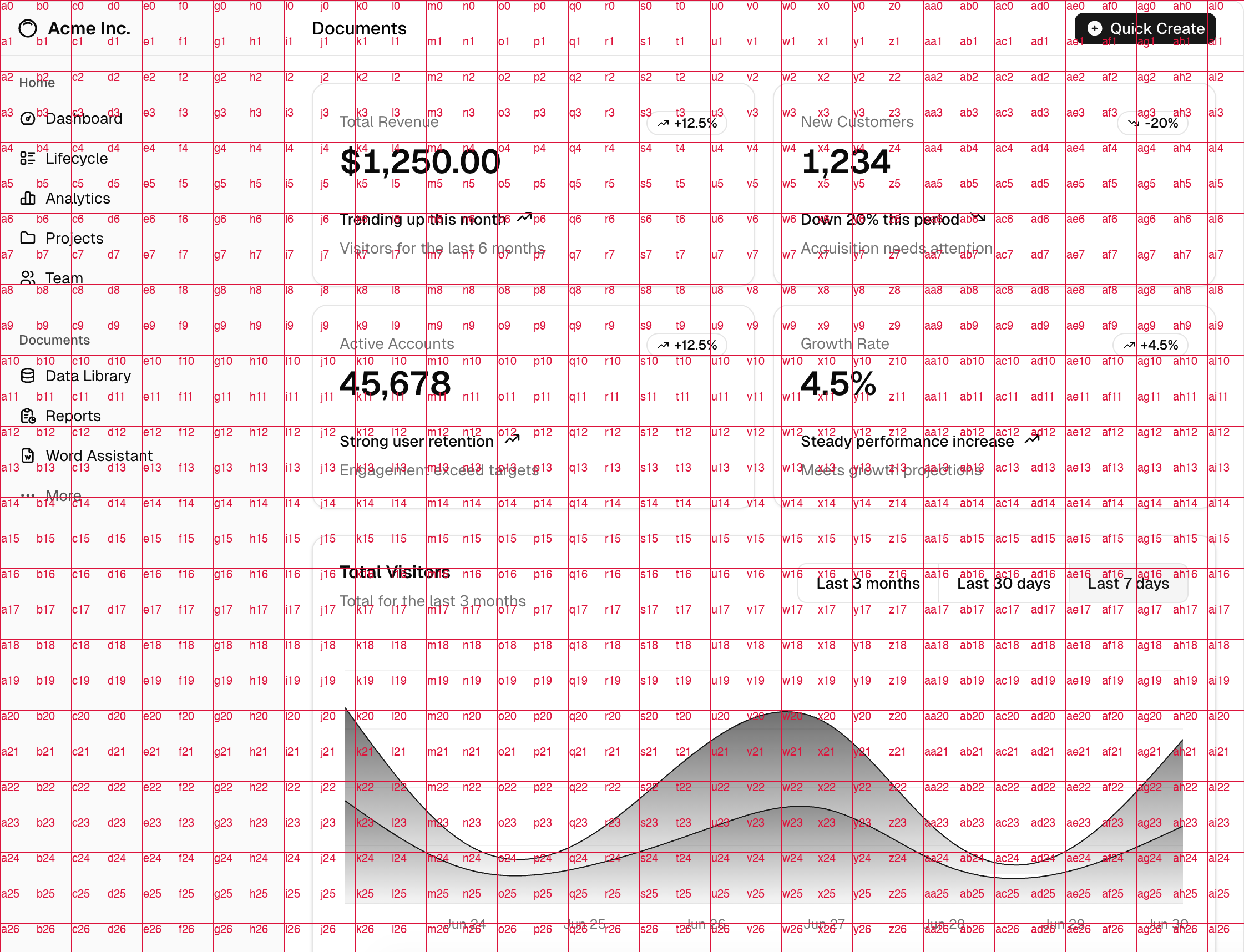}
	\caption{\textbf{Grid-based coordinate mapping approach.} A screenshot overlaid with a chess-style grid system where the LLM can reference elements by grid coordinates (for example, ``click 4x7''). While this simplifies coordinate communication, it becomes problematic with dense interfaces or small clickable elements.}
	\label{fig:dashboard-grid}
\end{figure}

Grid-based coordinate mapping attempts to simplify coordinate selection by overlaying a regular grid on top of the screenshot. The agent refers to grid cells (for example, ``click 4x7'') instead of specifying precise pixel coordinates. Figure~\ref{fig:dashboard-grid} illustrates this approach on a dashboard interface. The intended benefit is to provide a coarse but stable address space that is easier for the LLM to reason about than raw bounding boxes.

In practice, grid-based mapping performs poorly on the same cases that break pure vision. Small clickable elements require either extremely fine grids or multiple grid levels, both of which increase complexity. To reliably click 24px buttons, the grid must become so dense that labels and cell boundaries cover the underlying interface. The agent then reasons about approximate positions instead of concrete elements and frequently misses the intended targets. The overlay also hides text, icons, and micro-interactions that are important for understanding the page. After experimentation, this approach was abandoned in favor of architectures that treat screenshots as a supporting signal instead of the primary interaction surface.

\subsection{Hybrid vision and accessibility context}

This architecture benefits from combining accessibility tree snapshots with selective vision as paired approaches. Accessibility tree snapshots provide a compact, semantic view of the page: labels, roles, focus state, and validation messages. Vision adds the ability to see non-DOM content, verify spatial relationships, and interpret elements rendered on canvas or images. Together, they allow the agent to plan actions using a structured representation and use vision when that structure is missing or misleading.

From a performance perspective, accessibility tree snapshots are effective at giving the agent global context in a single request. An agent can see navigation, forms, dialogs, and error messages across the entire page without scrolling, and can plan a batch of tool calls that operate on multiple fields at once. This capability connects directly to the bulk action patterns in Section~\ref{sec:execution-layer}: while a human scrolls and clicks through each field sequentially, the agent can fill dozens of inputs or toggle multiple controls in a single bulk action. Vision supplements this by handling elements that never appear in the accessibility tree snapshot, such as charts, canvas-based editors, and game interfaces.

In this architecture, the execution layer automatically provides a fresh accessibility tree snapshot after each interaction, keeping the primary LLM synchronized with the current page state. When the agent encounters elements that are not exposed in the accessibility tree snapshot or require visual understanding, it can invoke the \texttt{take\_screenshot} tool. For cases that require precise element identification within visual content, the agent can delegate to a separate vision model with bounding box detection capabilities (such as Gemini 2.5 Pro\citep{google2025geminivision}). This vision model receives the screenshot, identifies clickable regions with their coordinates, and returns a structured representation similar to an accessibility tree snapshot (for example, \texttt{ref=5 "14 November"}). The primary LLM then interacts with these elements using the same ref-based tool calls it uses for accessibility tree elements. This delegation pattern keeps most interactions fast by operating on text-based snapshots, while still providing a fallback for visually rendered content that lacks semantic structure.

Hybrid designs also enable stronger safety boundaries. When agents operate primarily through accessibility tree snapshots, the execution layer can restrict actions based on semantic information: for example, by blocking clicks on elements whose labels include ``refund'' or ``delete'' unless explicit confirmation is present. Coordinate-based clicking provides few such guarantees. By defaulting to accessibility-based actions and using vision as a secondary signal for supplementary context, the system can apply programmatic constraints that are difficult to enforce when the agent interacts through coordinates alone.

Context management for browser agents spans both page perception and temporal memory. Pure vision approaches resemble how humans see the web but struggle with dense layouts, canvas-based applications, and safety. Pure accessibility tree snapshots provide compact, semantic representations but rely on WCAG-compliant markup and can miss dynamic behaviors or visual cues. Grid-based coordinate mapping offers little benefit in practice and complicates dense interfaces. Production experience, combined with evaluation results in Section~\ref{sec:production-validation}, supports a hybrid approach in which accessibility tree snapshots form the primary context for planning and tool calls, and vision is used selectively for non-accessible or highly visual tasks.

\section{Execution Layer}
\label{sec:execution-layer}

The execution layer links the large language model's reasoning and the browser's DOM. While the context management layer perceives the page state, the execution layer translates the agent's intent into concrete browser actions. This component must handle the complexity of modern web applications, where dynamic content, layout shifts, and state changes frequently desynchronize the agent's mental model from the actual page state.

\subsection{Element Reference System}

A fundamental challenge in browser automation is establishing a reliable reference system between the LLM and the DOM elements. While humans interact visually, browser agents often rely on structured text representations. The system must provide a mechanism for the LLM to reference specific elements in its tool calls.

The standard approach involves generating unique identifiers (refs) for interactive elements within the accessibility tree snapshot. The execution layer maintains a mapping between these generated refs and the actual DOM nodes.

\begin{figure}[H]
\centering
\begin{minipage}{\linewidth}
\begin{lstlisting}[language=Java]
const buttons = document.querySelectorAll('button');
const refMap = new Map();

buttons.forEach((button, index) => {
  const ref = `btn_${index}`;
  refMap.set(ref, button);
});

/* The agent references the element by its generated ref */
const element = refMap.get('btn_0');
element.click();
\end{lstlisting}
\end{minipage}
\caption{\textbf{Element reference mapping.} The execution layer assigns unique identifiers to DOM elements, allowing the LLM to target actions without needing complex CSS selectors.}
\label{fig:ref-mapping}
\end{figure}

\subsubsection{State Divergence and Versioning}

A critical issue with this reference system is state divergence. Modern web interfaces are highly dynamic; a "Cancel" button with \texttt{ref=10} in one snapshot might be replaced by a "Delete" button with the same ref in a subsequent render, or the element might simply disappear. If the agent operates on stale references, it risks executing unintended actions.

To address this, the system employs snapshot versioning. Each element reference includes a version identifier (e.g., \texttt{1:10}, where \texttt{1} is the snapshot version and \texttt{10} is the element ref). When the execution layer receives a tool call, it verifies that the requested version matches the current state. If the versions mismatch, the action fails safely, preventing interaction with incorrect elements.

Alternative approaches involve generating globally unique, multi-character IDs for every element across all snapshots. While this eliminates collision risks, it increases token consumption. The versioned integer approach offers a balance between safety and token efficiency.

\subsection{Tool Definitions}
\label{sec:tools}

The browser agent interacts with the web page through a defined set of tools. These primitives cover the range of human browser interactions, from basic clicks to complex state management.

\subsubsection{Interaction Tools}

\texttt{click(ref, doubleClick?, rightClick?, holdMs?)}: Performs a click action on the specified element. While standard clicks make up the majority of interactions, the tool supports double-clicks and right-clicks to handle context menus and specialized interfaces, although these are less frequent in standard web navigation.

\texttt{type(ref, text, shouldClear?)}: Enters text into input fields. Optionally can clear existing content before typing.

\texttt{hover(ref)}: Triggers hover states, which is essential for revealing dropdown menus, tooltips, and other on-hover UI elements.

\texttt{press\_key(key)}: Simulates keyboard events. This tool is useful for accessibility-first interfaces and failure recovery. When standard element interaction fails, agents often resort to keyboard navigation (e.g., \texttt{Tab} to focus, \texttt{Enter} to activate) or shortcuts (e.g., \texttt{Cmd+C} / \texttt{Cmd+V} for clipboard operations, arrow keys for document editing).

\texttt{select\_option(ref, values)}: Selects values in standard \texttt{<select>} elements and accessible listboxes.

\texttt{upload\_file(ref, filePaths)}: Handles file input interactions by attaching specified files to the form element.

\texttt{drag(startRef, endRef)}: Executes drag-and-drop operations between two elements, necessary for kanban boards, sliders, and ordering interfaces.

\texttt{pan(ref?, deltaX, deltaY)}: Controls viewport or container scrolling, enabling interaction with map interfaces and canvas elements.

\texttt{focus(ref)}: Explicitly sets focus to an element and scrolls it into view, often used as a precursor to keyboard events.

\subsubsection{State and Navigation Tools}

\texttt{wait\_for(time?, textToWait?, textGone?)}: Pauses execution until a condition is met. This tool is critical for handling asynchronous state changes, such as waiting for a loading spinner to disappear or a success message to appear. Proper use of waiting mechanisms allows the agent to synchronize with the application state effectively.

\texttt{handle\_dialog(accept, promptText?)}: Manages native browser dialogs (alerts, confirms, prompts) that would otherwise block execution.

\texttt{navigate(url)}: Loads a specific URL.

\texttt{navigate\_back()}: Simulates the browser's back button.

\texttt{browser\_tabs(action, url?, tabId?)}: Manages browser tabs, including creating new tabs, switching focus, and closing tabs. This enables multi-tab workflows such as cross-referencing data between pages.

\texttt{snapshot(ref?, mediaType?, startRef?, endRef?)}: Requests a fresh accessibility tree snapshot. The \texttt{mediaType} parameter allows the agent to request a "print" view, which can simplify content extraction by removing advertisements and navigation elements, similar to a browser's reader mode.

\texttt{take\_screenshot(ref?, fullPage?)}: Captures a visual screenshot of the viewport or a specific element. This is primarily used when the accessibility tree snapshot is insufficient for understanding the page state.

\subsection{Bulk Actions}

Sequential execution of actions introduces significant latency, particularly in form-filling tasks. To address this, the architecture supports bulk actions, enabling the agent to dispatch multiple actions in a single tool call. Modern LLMs can generate JSON structures representing a sequence of independent actions.

\begin{figure}[H]
\centering
\begin{minipage}{\linewidth}
\begin{lstlisting}[language=Java]
bulkActions([
  { type: "type", ref: 1, text: "hello" },
  { type: "type", ref: 2, text: "world" },
  { type: "select_option", ref: 3, values: ["USA"] }
])
\end{lstlisting}
\end{minipage}
\caption{\textbf{Bulk action structure.} Grouping independent interactions reduces network round-trips and LLM inference overhead.}
\label{fig:bulk-actions}
\end{figure}

This batching capability significantly improves performance for tasks that do not require intermediate verification, such as populating multiple fields in a form.

\begin{figure}[H]
	\centering
	\includegraphics[width=\linewidth, alt={Chart comparing performance of bulk vs sequential form filling}]{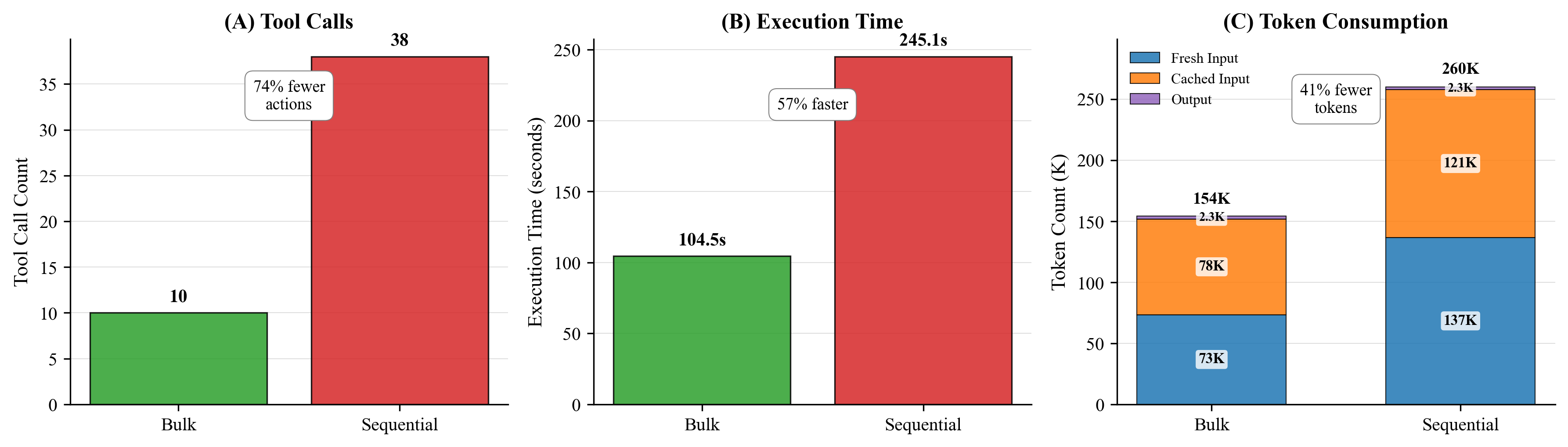}
	\caption{\textbf{Performance comparison of bulk vs sequential action approaches on form filling.} Tested on a form with 28 fields (text inputs, dropdowns, radio buttons) using GPT-5.1. The bulk approach batches multiple field interactions into single tool calls, while the sequential approach processes each field individually. Results show the bulk approach achieves 74\% fewer tool calls (10 vs 38), executes 57\% faster (104.5s vs 245.1s), and consumes 41\% fewer tokens (154K vs 260K total).}
	\label{fig:bulk-sequential-comparison}
\end{figure}

As demonstrated in Figure \ref{fig:bulk-sequential-comparison}, the bulk action approach improves efficiency, reducing both execution time and token costs compared to sequential processing.

\subsection{Error Handling and Adaptation}

Browser agents frequently encounter execution failures due to stale references, obscured elements, or non-standard UI implementations. For instance, a custom dropdown might not respond to the standard \texttt{select\_option} tool, requiring the agent to simulate a click and a following selection.

Effective error handling requires clear feedback from the execution layer. When an action fails, the system provides a descriptive error message (e.g., "Element not clickable," "Ref detached from the DOM"). This feedback, combined with a fresh page snapshot, enables the agent to diagnose the failure and attempt a recovery strategy.

The system prompt explicitly instructs the agent to adapt its approach upon failure instead of blindly retrying the same action.

\begin{figure}[H]
\centering
\begin{minipage}{\linewidth}
\begin{lstlisting}
<failure_adaptation>
When an action fails:
1. Never retry the same action twice
2. Analyze why it failed (error message, page state, element availability)
3. Try fundamentally different approach:
   - Different element/interaction method
   - Different tool or sequence
4. If second approach fails, reconsider the goal itself
</failure_adaptation>
\end{lstlisting}
\end{minipage}
\caption{\textbf{Failure adaptation protocol.} System instructions guide the agent to attempt alternative interaction methods when primary actions fail.}
\label{fig:failure-adaptation}
\end{figure}

This adaptive behavior is essential for autonomous operation, allowing the agent to navigate around the inconsistencies and edge cases built into modern web interfaces.

\section{Context \& History Management}
\label{sec:memory-management}

Context management is a determining factor in the reliability and cost-efficiency of browser agents. While LLMs have the capability to process large amounts of information, the presence of irrelevant data in the context window increases latency, cost, and the probability of hallucination. A browser agent must maintain sufficient information to understand the current state and past actions while discarding noise that masks the objective.

Due to the nature of MCPs, standard implementations such as Playwright MCP and Chrome DevTools MCP retain the full conversation history, including all intermediate page snapshots. This approach results in linear growth of token consumption with each action. For example, if an accessibility tree snapshot consumes 10,000 tokens, a workflow requiring 20 actions would accumulate over 200,000 tokens in the context window. This accumulation makes long-running tasks prohibitively expensive and degrades LLM performance as the context length increases.

To address these challenges, the architecture employs a three-tiered strategy: retaining only the most recent accessibility tree snapshot, applying intelligent trimming to that snapshot, and compressing the conversation history.

\subsection{Snapshot Management}

The system retains only the most recent accessibility tree snapshot in the system prompt. Unlike chat-based interfaces that preserve the entire dialogue, the browser agent treats the page state as temporary. When the agent navigates to a new page or performs an action that updates the view, the previous snapshot is discarded. This approach ensures that the token consumption for the page state remains constant regardless of the task duration.

However, even a single snapshot can exceed the optimal context window for efficient reasoning. Complex e-commerce pages or dashboards often contain thousands of elements. To manage this, the system enforces a maximum snapshot size (e.g., 50,000 characters). If the raw accessibility tree exceeds this limit, it undergoes intelligent trimming.

\begin{figure}[H]
\centering
\begin{minipage}{\linewidth}
\begin{lstlisting}
ref=1000 button "Hello World"
...
[refs 1001-5000 trimmed]
\end{lstlisting}
\end{minipage}
\caption{\textbf{Snapshot truncation indicator.} When a snapshot is truncated, the agent receives an explicit indicator of the missing range. The \texttt{snapshot} tool allows the agent to request specific ranges (e.g., \texttt{startRef=1001}, \texttt{endRef=2000}) if the trimmed content is required.}
\label{fig:snapshot-trimming}
\end{figure}

\subsection{Intelligent Trimming}

Simple character-based truncation is insufficient because it may arbitrarily cut critical interactive elements while retaining repetitive content. To solve this, the system employs a lightweight model (e.g., Gemini 2.5 Flash Lite) to perform intelligent trimming before the snapshot reaches the primary LLM.

The process involves sending the full accessibility tree snapshot along with the current conversation history to the lightweight model. This model identifies the sections of the page that are relevant to the user's current objective and returns a list of element reference ranges to retain. The system then constructs a filtered snapshot containing only the identified ranges.

The prompt for the lightweight model instructs it to preserve all interactive elements (navigation, forms, buttons, modals) while aggressively summarizing repetitive content such as long lists of product cards. For example, in a list of 50 products, the trimmer might retain the first 5 items to establish the pattern and trim the remainder, while keeping the pagination controls visible.

\begin{figure}[H]
\centering
\begin{minipage}{\linewidth}
\begin{lstlisting}
const prompt = `You are a snapshot analyzer for a browser agent.

Your task: Identify which parts of the accessibility tree snapshot are relevant 
to the user's current request. Aggressively trim repetitive content while 
preserving ALL interactive elements.

<conversation_history>
<user_request>Complete the checkout form with my shipping information</user_request>
<step>Navigated to checkout page, page loaded successfully</step>
<step>Found checkout form with fields: name, email, address, city</step>
</conversation_history>

RULES:
- KEEP all navigation, forms, buttons, modals, dialogs
- TRIM repetitive lists to first 5 items
- Keep 30-40 refs context around important elements

SNAPSHOT (${totalRefs} refs):
${snapshot}

Return: [{start: X, end: Y}, ...]`;

const response = await llmCall({
  model: 'gemini-2.5-flash-lite',
  prompt
});
// returns [{start: 0, end: 50}, {start: 200, end: 250}]
\end{lstlisting}
\end{minipage}
\caption{\textbf{Intelligent trimming logic.} A lightweight model analyzes the page structure and user intent to select relevant element ranges, significantly reducing the token load for the primary LLM.}
\label{fig:intelligent-trimming-code}
\end{figure}

\begin{figure}[H]
	\centering
	\includegraphics[width=\linewidth, alt={Chart comparing cost with and without intelligent trimming}]{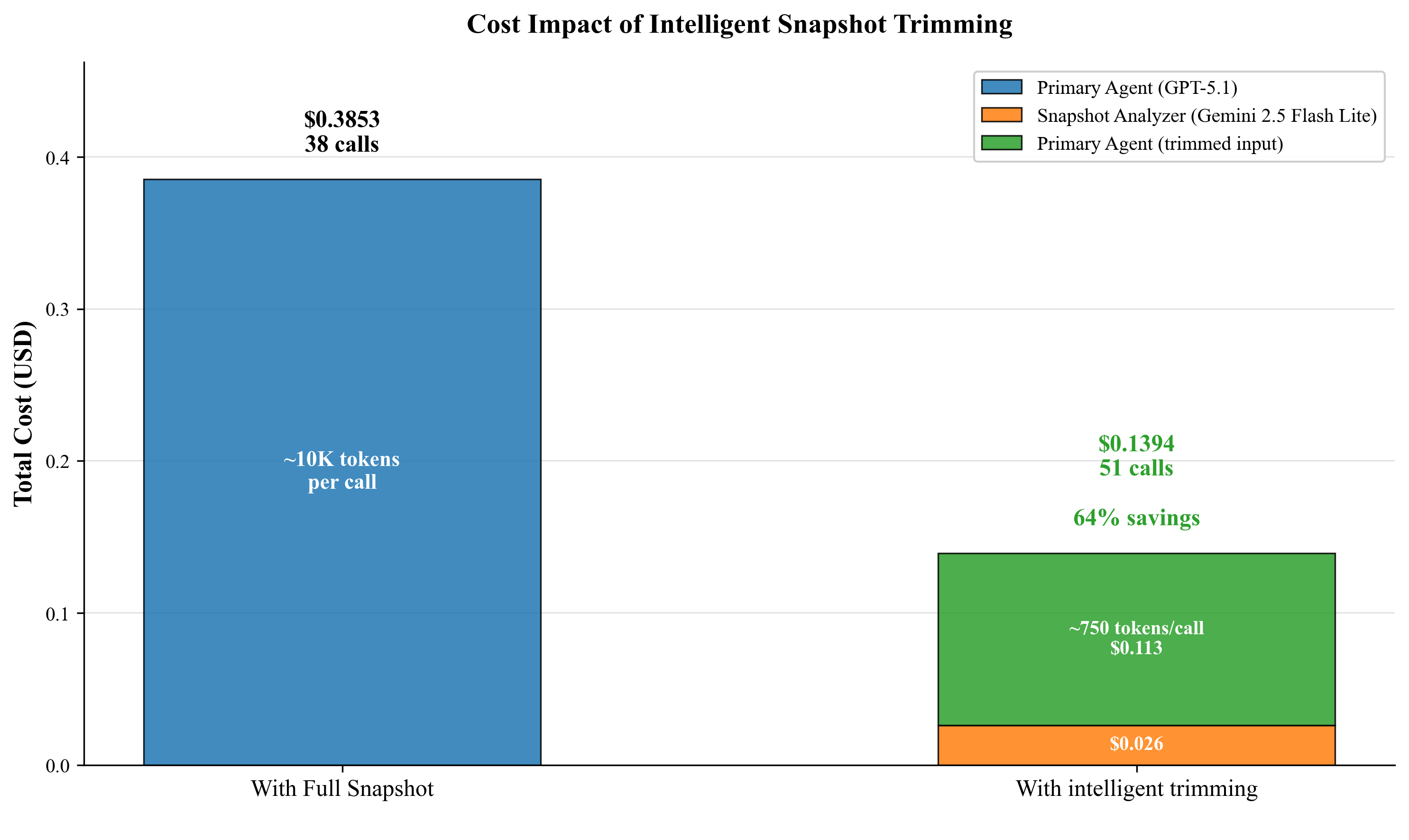}
	\caption{\textbf{Intelligent Trimming Cost Analysis.} Comparison of two approaches over a 38-51 action task. Without trimming, the primary LLM receives full snapshots on every call. With intelligent trimming, a lightweight model filters the snapshot, reducing the input size for the primary LLM from ~10,000 tokens to 500-1,000 tokens per step. Despite a 34\% increase in tool calls due to occasional re-requests, the total cost decreases by 57\%.}
	\label{fig:intelligent-trimming}
\end{figure}

This approach significantly reduces costs, though the magnitude of savings is directly correlated with the complexity of the target page. Larger snapshots produce greater savings, as the lightweight model filters out a higher volume of irrelevant tokens. As shown in Figure~\ref{fig:intelligent-trimming}, using a lightweight model to filter content reduces the total cost by approximately 57\% for long-running tasks on pages with 8,000-12,000 token snapshots, even accounting for the additional tool calls required if the primary agent requests missing content.

\subsection{Conversation History Compression}

To prevent the context from growing indefinitely, the system compresses the conversation history. Instead of retaining the full log of messages, tool calls, and outputs, the system summarizes completed steps into a concise log.

Each tool call includes a \texttt{memory} parameter, where the agent explicitly summarizes the outcome of the action and updates its internal state. This self-generated summary serves as the persistent memory for the agent. This approach was originally introduced by Browser Use~\cite{browseruse2024} and has proven effective for maintaining context across long-running tasks.

\begin{figure}[H]
\centering
\begin{minipage}{\linewidth}
\begin{lstlisting}[language=Java]
click({
  ref: 50,
  evaluation_previous_goal: "Successfully clicked submit button",
  memory: "Form submitted, confirmation page loaded with order #12345",
  next_goal: "Email order number to email@example.com"
})
\end{lstlisting}
\end{minipage}
\caption{\textbf{Self-correction and memory update.} The agent uses the \texttt{memory} parameter to document its progress, creating a compressed history log that persists across steps.}
\label{fig:memory-parameter}
\end{figure}

The system maintains a rolling buffer of the most recent actions (e.g., the last 40-50 steps) in their full detail, while older steps are summarized or discarded. This ensures that the agent retains immediate context for error recovery while maintaining a global view of the task progress through the compressed memory log.

\begin{figure}[H]
\centering
\begin{minipage}{\linewidth}
\begin{lstlisting}
<initial_user_request>fill the shipping form</initial_user_request>

<step>
Successfully found the form
Form has 5 fields: name, address, city, zip, country
Fill the name field first
click(ref=42), type(ref=42, "John Doe")
</step>

<follow_up_user_request> use my home address </follow_up_user_request>

<step>
Name field filled successfully
User wants home address which is 123 Main St
Fill address field next
type(ref=45, "123 Main St"), type(ref=46, "New York")
</step>
\end{lstlisting}
\end{minipage}
\caption{\textbf{Compressed conversation history.} The system presents a summarized log of past actions and user interactions, allowing the agent to track progress without processing the full raw history.}
\label{fig:compressed-history}
\end{figure}

This compression is critical for multi-tab workflows. When an agent switches between tabs, it loses access to the previous page's snapshot. The \texttt{memory} field ensures that information extracted from one tab (e.g., "Order ID is 12345") is explicitly recorded and available when the agent switches to another tab to enter that data.

\begin{figure}[H]
	\centering
	\includegraphics[width=\linewidth, alt={Line graph showing linear token growth vs constant compressed history}]{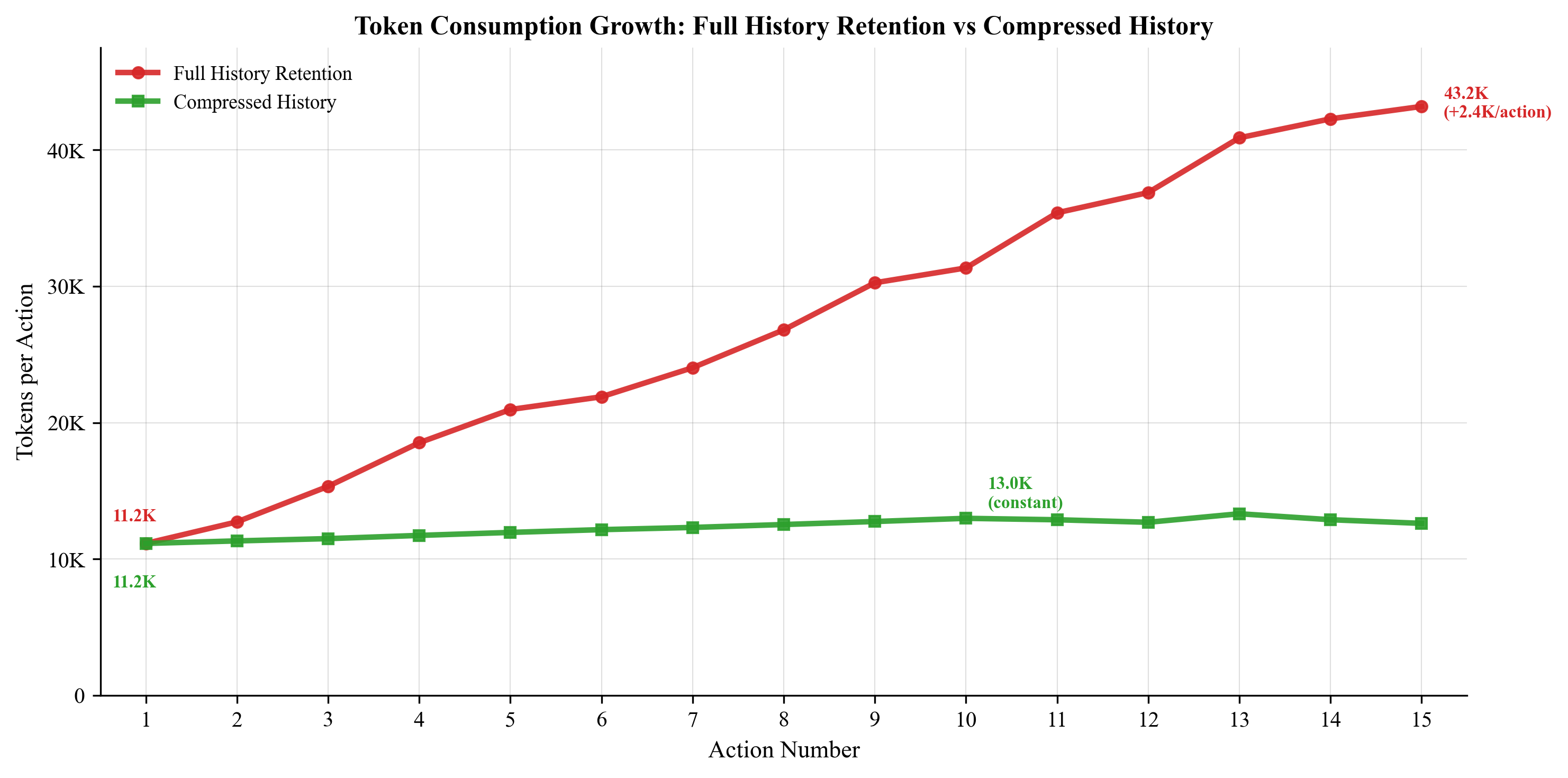}
	\caption{\textbf{Token consumption: Full History vs. Compressed History.} Full history retention leads to linear growth, reaching over 43,000 tokens after 15 actions. The compressed history approach stabilizes around 12,600 tokens, maintaining constant performance and cost regardless of task length.}
	\label{fig:token-growth-comparison}
\end{figure}

By combining single-snapshot retention, intelligent trimming, and history compression, the architecture maintains a stable token consumption profile. As illustrated in Figure~\ref{fig:token-growth-comparison}, this approach prevents the linear token growth observed in naive implementations, enabling the agent to execute tasks of unlimited length within a fixed cost and latency budget.

\section{Prompt Engineering \& Optimization}
\label{sec:prompt-engineering}

The system prompt defines the browser agent's role, capabilities, and operational rules, including its core mission, behavior patterns, failure recovery strategies, and task completion criteria. Effective prompt engineering for browser agents requires addressing the unique constraints of web automation: latency, token costs, and the need for temporal awareness.

\subsection{System prompt structure}

The system prompt establishes the agent's fundamental behavior. It provides critical operational context that the LLM lacks, specifically regarding the time, cost of actions and the need for persistent error recovery.

\begin{figure}[H]
\centering
\begin{minipage}{\linewidth}
\begin{lstlisting}
You are an AI browser automation agent designed to understand user requests 
and execute them autonomously using available tools.

<agent_behavior>
- Goal-oriented: Complete tasks efficiently and persistently
- Adaptive: Try different approaches when actions fail
- Autonomous: Work independently until task completion
- Time-aware: Each tool call takes ~3-5s; batch actions aggressively to minimize latency
</agent_behavior>

<tools>
click: {
  ref: number - Element reference from accessibility tree snapshot
  evaluation_previous_goal: string - Assessment of last action result
  memory: string - Key progress and information for next steps
  next_goal: string - Immediate next action to take
  doubleClick?: boolean - Perform double-click
  rightClick?: boolean - Open context menu
}

type: {
  ref: number - Element reference to type into
  text: string - Text to enter
  ...
}
...
</tools>
\end{lstlisting}
\end{minipage}
\caption{\textbf{System prompt configuration.} The prompt explicitly defines the agent's operational constraints, including time awareness and the structure of available tools.}
\label{fig:system-prompt}
\end{figure}

Time awareness is critical because LLMs lack the temporal perception that humans naturally have. The agent must understand the current time, the duration of each action, and the cumulative time spent on a task. Production observations indicate that without this context, agents often pursue inefficient, step-by-step strategies that become non-viable in time-sensitive scenarios. When provided with explicit temporal context (e.g., "Each tool call takes ~3-5s"), agents optimize their execution strategy by batching actions more aggressively and prioritizing faster approaches. In timed scenarios, such as assessments or auction interfaces, this awareness allows the agent to restructure its plan to complete all necessary interactions within the deadline.

\subsection{Caching strategy}

Dynamic context, including the page snapshot, browser tabs, selected text, and compressed history, changes with each action. To optimize performance and cost, the system injects this context separately from the static system prompt. This separation enables effective prefix caching \citep{openai2024caching}, as the static portion of the prompt remains constant across requests.

An effective caching strategy orders content by change frequency, placing the most stable elements at the beginning of the context window:

\begin{enumerate}
\item \textbf{System prompt:} Static instructions that never change.
\item \textbf{Session context:} User locale, timezone, and custom instructions (changes per session).
\item \textbf{Tab state:} List of open browser tabs (changes on navigation).
\item \textbf{Conversation history:} Compressed log of past actions (appends each step).
\item \textbf{Current page snapshot:} The accessibility tree snapshot (changes every action).
\end{enumerate}

The accessibility tree snapshot is positioned last as the most frequently changing component. This ordering ensures that all preceding content can be cached if it remains unchanged between requests.

\subsection{Cost optimization}

This caching architecture significantly reduces the operational cost of long-running workflows. For a workflow consisting of 100 requests, the cost difference is substantial. Without caching, a 20,000-token system prompt would be processed fully for every step, resulting in 2 million processed tokens (\$2.50 at GPT-5.1 rates). With caching, the system pays the full input cost only for the first request. Later requests treat the static prefix as cached input, which is typically priced at a fraction of the standard rate (e.g., \$0.13/1M tokens vs \$1.25/1M tokens for GPT-5.1). This approach results in cost reductions of approximately 89\% for extended sessions, making autonomous operation economically viable.

\section{Security \& Safety}
\label{sec:security}

Security remains the most significant barrier preventing browser agents from autonomous production operation. The attack surface for these systems differs fundamentally from traditional web security models. While conventional applications rely on deterministic code-level protections, browser agents operate on natural language processing, introducing vulnerabilities that cannot be mitigated through standard input validation or sandboxing alone.

\subsection{The Fundamental Problem}

Traditional web security relies on code-level protections such as input validation, sandboxing, and the same-origin policy to isolate untrusted content. Browser agents, however, introduce a new attack vector where the vulnerability lies within the LLM's language processing capabilities. When an agent processes web content, it cannot reliably distinguish between legitimate user commands and malicious instructions embedded in the page. This phenomenon, known as prompt injection, allows attackers to override the agent's instructions by hiding commands in the text or metadata of a website or the dynamic content of trusted websites \citep{liu2024prompt}.

Current LLMs lack reliable solutions for this problem. Unlike SQL injection or cross-site scripting (XSS), which target strict syntax parsers, prompt injection targets the semantic reasoning layer of the LLM. As a result, purely prompt-based defenses typically fail to provide the absolute guarantees required for secure operation.

\subsection{Production Observations}

As detailed in Section~\ref{sec:prompt-injection}, public security analyses of AI browsers and browser extensions show that prompt injection attacks remain highly effective despite mitigation attempts. Studies involving systems such as Perplexity Comet indicate that attackers can execute cross-domain actions, such as accessing email or banking sessions, by embedding invisible text in a web page \citep{brave2025comet}.

Production experience confirms these risks. Once a browser agent receives broad tool access over a user's authenticated sessions, preventing untrusted content from triggering sensitive actions through LLM judgment alone becomes statistically unlikely. The failure rate of prompt-based defenses, even if only 1\%, represents an unacceptable risk when the agent has access to financial or communication platforms.

\subsection{Limitations of Traditional Defenses}

Standard security models for agents often rely on two approaches: permission-based controls and classifier-based detection. Both show major limitations in production:

\textbf{Permission-based controls} require the user to confirm actions or approve site access. While this provides a first line of defense, it creates "permission overload" leading users to habitually approve requests without inspection. Furthermore, users may not understand the implications of granting an agent access to a complex DOM environment where hidden elements can trigger actions.

\textbf{Classifier-based detection} employs classifier models to scan inputs for malicious patterns. While the attacker cannot directly communicate with the classifier, the classifier still processes the same untrusted web content. This means attackers can embed prompt injections that either appear safe to the classifier or exploit the classifier itself. This approach raises the bar for attackers but does not eliminate the vulnerability. For production systems handling sensitive data, layering one vulnerable model on top of another provides no reliable security boundary.

\subsection{Architectural Countermeasures}

Given the fundamental vulnerability of LLMs to prompt injection, secure autonomous operation requires removing security decisions from the LLM's domain. Production experience suggests that safety must be enforced through deterministic, programmatic constraints instead of probabilistic reasoning.

\subsubsection{Deterministic Safety Boundaries}
Safety policies should be enforced by the execution layer code, not the LLM. By using the structured nature of accessibility tree snapshots, the system can block interactions with sensitive elements based on deterministic rules.

\begin{figure}[H]
\centering
\begin{minipage}{\linewidth}
\begin{lstlisting}[language=Java]
const element = getElementByRef({ ref });
const sensitiveKeywords = ["refund", "delete", "transfer", "password"];

// Programmatic check enforcing confirmation for sensitive actions
if (sensitiveKeywords.some(keyword => 
    element.text.toLowerCase().includes(keyword)
  )) {
  if (!requireConfirmation) {
    throw new Error(`Action on '${element.text}' requires explicit user confirmation`);
  }
}
\end{lstlisting}
\end{minipage}
\caption{\textbf{Programmatic safety check.} The execution layer inspects the target element's properties before executing an action, enforcing safety policies that the LLM cannot override.}
\label{fig:safety-check}
\end{figure}

\subsubsection{Domain Allowlisting}
General-purpose browsing presents an unbounded attack surface. To address this, production agents should operate under strict domain allowlisting. If an agent is designed to "extract CRM data and update Salesforce," the architecture must physically restrict network access to only the CRM domain and Salesforce. This prevents the agent from visiting attacker-controlled sites or exfiltrating data to third-party servers, regardless of any injected instructions it might encounter.

\subsubsection{Action Restriction}
The execution layer can programmatically restrict dangerous actions. As shown in the code example above, the system blocks clicks on elements containing sensitive keywords such as "refund," "delete," "transfer," or "password" unless explicit user confirmation is provided. This ensures that even if the LLM is compromised through prompt injection, the execution layer prevents irreversible actions from being executed.

\subsubsection{Specialization as Defense}
The most effective defense is specialization. Instead of a single "general intelligence" agent with access to email, banking, and documents, the architecture should use specialized agents with least-privilege scopes. A "LinkedIn Researcher" agent requires access only to LinkedIn and public search, with no ability to access email or internal file systems. This compartmentalization limits the blast radius of any potential compromise. Section~\ref{sec:agent-specialization} discusses concrete examples of specialized agent types and their operational constraints.

\subsection{Summary}
While browser agents offer significant productivity benefits, security fundamentals remain unsolved at the LLM level. Reducing risk by 90\% is insufficient for autonomous systems handling sensitive data. Secure production operation requires applying proven software security principles, such as programmatic enforcement, least privilege, and sandboxing, to the agent architecture. Production agents must be specialized, constrained by code, and skeptical by design, relying on architectural boundaries over LLM intelligence for safety.

\section{Production Validation}
\label{sec:production-validation}

Taking a browser agent to production requires measuring factors beyond simple task completion: operational costs, reliability at scale, and the safety of autonomous execution. While previous sections addressed architecture and security mechanisms, this section presents concrete data from production operation and benchmark evaluation.

\subsection{Cost Analysis}
\label{sec:cost-analysis}

This analysis examines real-world costs using the WebGames Shopping Challenge benchmark \citep{webgames2025}, specifically the "Cheapest product addition challenge." This task requires an agent to navigate multiple product pages, systematically compare prices, identify the lowest-priced items, and add them to a shopping cart. The challenge demands advanced reasoning capabilities: the agent must maintain a mental model of prices, avoid suboptimal selections, and efficiently navigate paginated listings without redundant steps.

\begin{figure}[H]
	\centering
	\includegraphics[width=0.95\linewidth, alt={Chart showing cost breakdown: input, output, and cached tokens}]{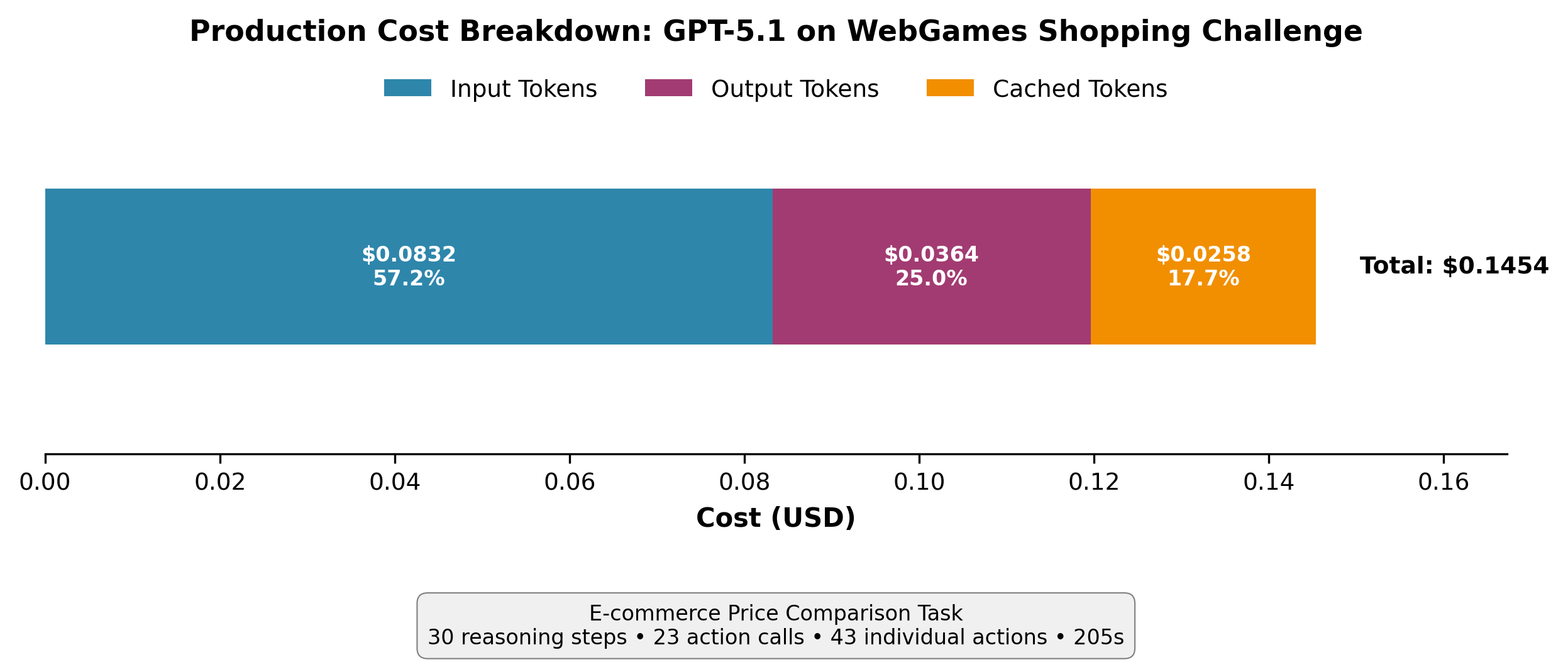}
	\caption{\textbf{Production cost breakdown for e-commerce price comparison task.} Total cost: \$0.1454 for 30 reasoning steps, 23 tool calls, and 43 individual actions over 205 seconds. Total tokens: 268,743 (265,104 input including 198,528 cached, 3,639 output). Cost breakdown: non-cached input \$0.0832 (57.2\%), output \$0.0364 (25.0\%), cached input \$0.0258 (17.7\%). Per-step average: \$0.0048.}
	\label{fig:webgames-cost}
\end{figure}

Evaluation using GPT-5.1 on the "Cheapest product addition challenge". The agent successfully completed the multi-step price comparison and checkout workflow in 3.4 minutes with a total cost of \$0.1454. As detailed in Figure~\ref{fig:webgames-cost}, caching played a major role in cost control, with nearly 75\% of input tokens served from cache.

\begin{figure}[H]
	\centering
	\includegraphics[width=\linewidth, alt={Charts of total token distribution and per-step metrics}]{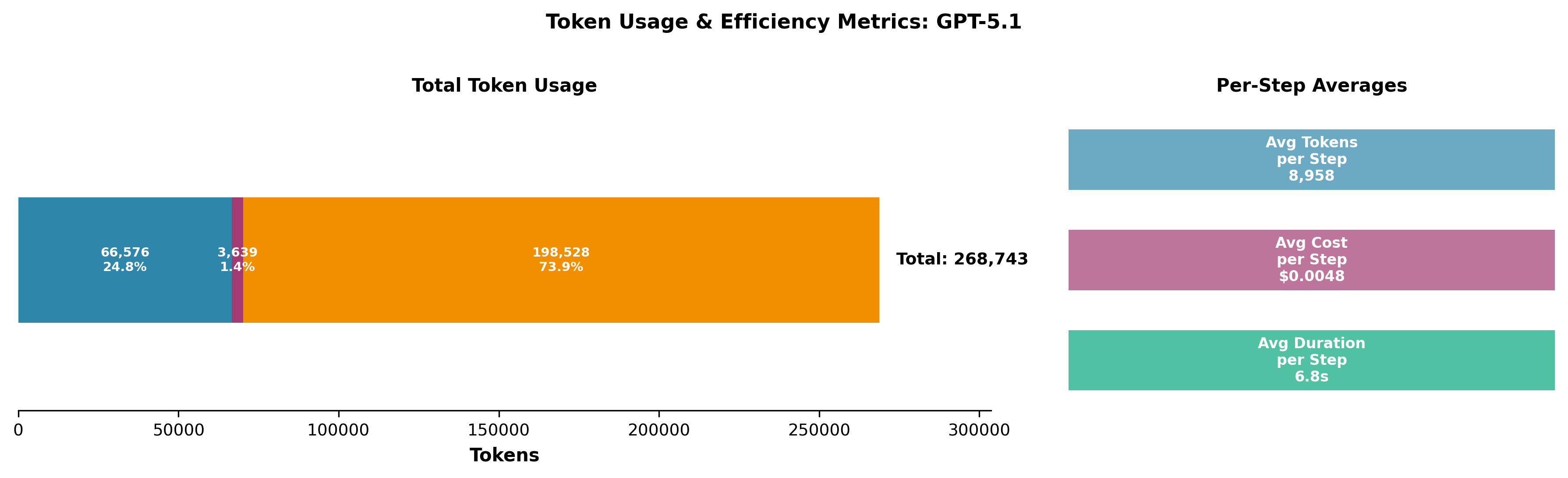}
	\caption{\textbf{Token usage and per-step metrics.} Left: Token distribution across 268,743 total tokens. Right: Per-step averages for capacity planning: 8,958 tokens/step, \$0.0048/step, 6.8s/step.}
	\label{fig:webgames-tokens}
\end{figure}

Figure~\ref{fig:webgames-tokens} illustrates token distribution and per-step efficiency metrics. The high cache ratio (74.9\% of input tokens) is typical for reasoning-heavy agents that process extensive context (page snapshots, price comparisons) but produce concise tool calls. Per-step averages indicate a predictable capacity planning baseline: 8,958 tokens, \$0.0048, and 6.8 seconds per step.

\begin{figure}[H]
	\centering
	\includegraphics[width=\linewidth, alt={Timeline bar chart of browser action execution durations}]{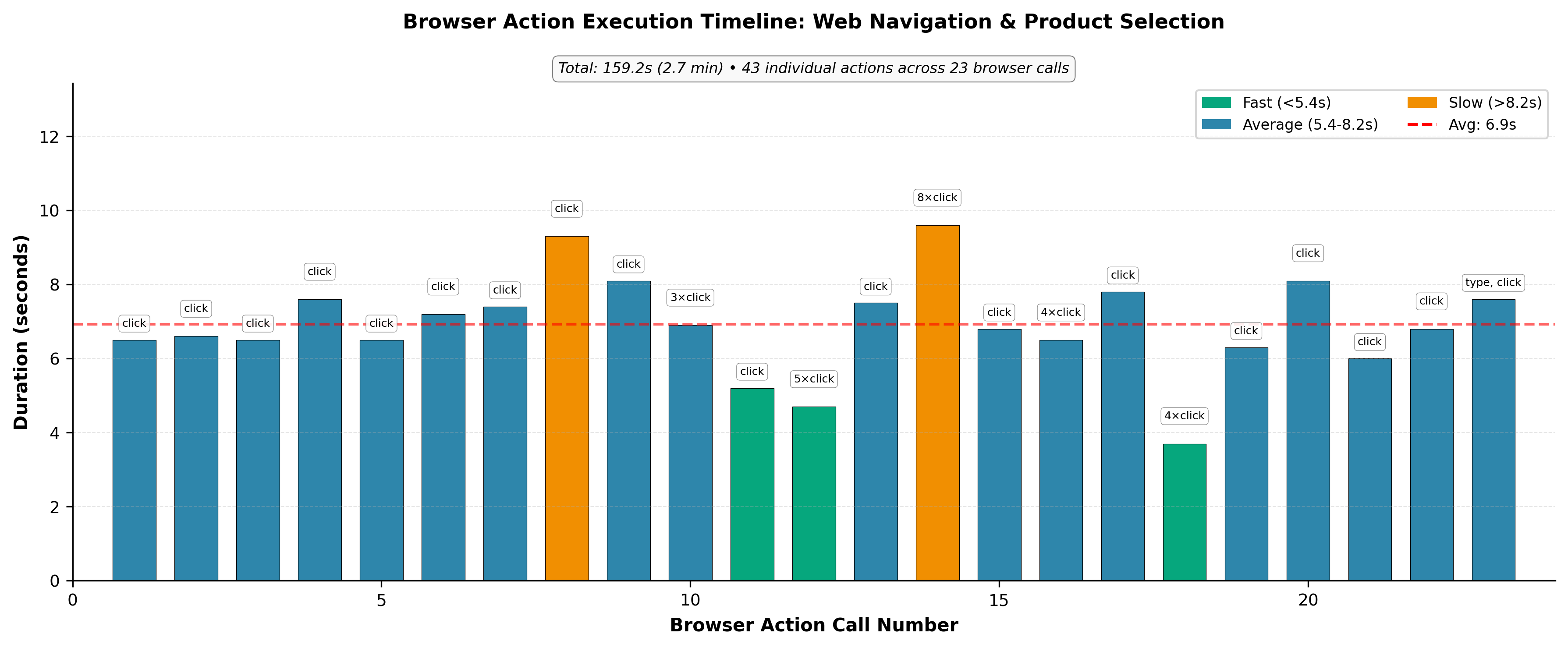}
	\caption{\textbf{Browser action execution timeline.} Each bar represents one of 23 tool calls, color-coded by speed: green (<5.4s), blue (5.4-8.2s), orange (>8.2s). Labels show action types executed (e.g., \texttt{click}, \texttt{3x click}, \texttt{type}). The 43 individual actions are distributed across these 23 calls, with some calls batching multiple actions for efficiency.}
	\label{fig:webgames-duration}
\end{figure}

Figure~\ref{fig:webgames-duration} shows execution timing for the 23 tool calls. Execution time varies from 3.7 to 9.6 seconds (average 6.8s). The workflow structure reveals optimization: early steps perform sequential page navigation (single clicks), while later steps use bulk actions (batching multiple actions) after the agent identifies optimal products.

\subsection{WebGames Benchmark Results}
\label{sec:benchmark-results}

Evaluation on the WebGames benchmark \citep{webgames2025}, which contains 53 challenges designed to test web agent capabilities across perception, reasoning, planning, and tool use, provides a standardized measure of performance. The benchmark is calibrated against human performance, with human subjects achieving a 95.7\% success rate.

The agent completed 45 out of 53 challenges successfully, achieving an approximately 85\% success rate. For comparison, the WebGames repository reports that the best-performing browser agent tested by the benchmark authors (Gemini 2.5 Pro with the Browser Use framework) achieved approximately 50\% success rate. The performance gap reflects the combination of the primary model's reasoning capabilities, comprehensive browser tooling matching human interaction capabilities (Section~\ref{sec:tools}), and the architectural patterns discussed in this paper.

\subsubsection{Failure Analysis}

Eight challenges remained incomplete, falling into three categories that highlight current technical limitations:

\textbf{Advanced vision requirements (5 challenges):} Tasks such as "Slider Symphony," "Color Harmony," and "Pixel Copy" require pixel-level spatial precision, subtle color differentiation, or exact visual pattern replication. These challenges specifically test pure vision capabilities of the model.

\textbf{Real-time interaction (2 challenges):} "Brick Buster" and "Frog Crossing" are arcade-style games requiring sub-second reaction times. Agent latency of 3-5 seconds per action makes these architecturally incompatible with current LLM inference speeds.

\textbf{Precision control (1 challenge):} "Block Stack" requires directional cursor movement with precise speed control for physics-based stacking, a modality not supported by discrete tool calls.

\begin{figure}[H]
	\centering
	\includegraphics[width=\linewidth, alt={Bar chart comparing success rates: Human 95.7\%, This Work 85\%, Prior AI 50\%}]{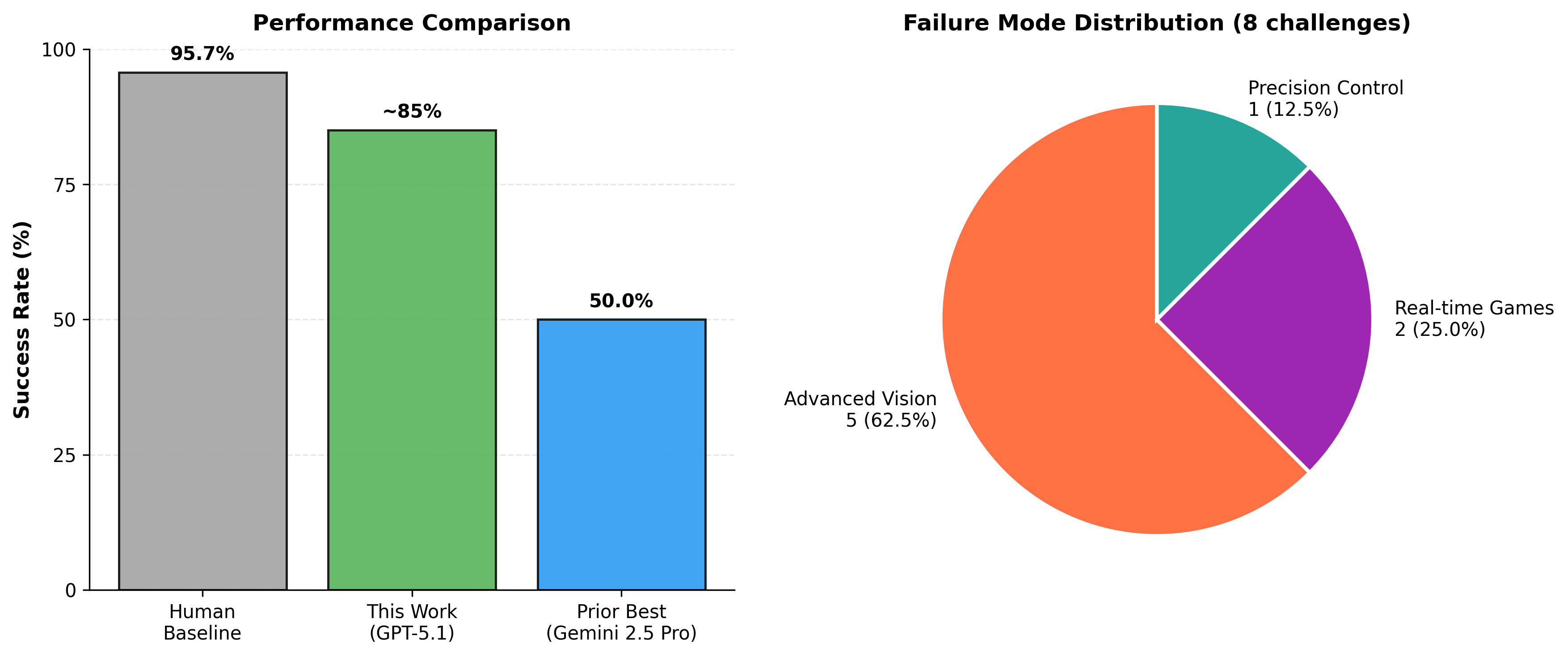}
	\caption{\textbf{WebGames benchmark evaluation.} (A) Performance comparison: approximately 85\% success compared to ~50\% reported for prior browser agents and 95.7\% human baseline. (B) The 8 unsuccessful challenges cluster into categories reflecting current latency and vision limitations instead of reasoning failures.}
	\label{fig:webgames-benchmark}
\end{figure}

\subsection{Operational Autonomy and Specialization}
\label{sec:agent-specialization}

A central challenge in production operation is balancing autonomy with safety. Requiring user confirmation for critical actions reduces risk but limits utility. Allowing full autonomy increases efficiency but introduces the risk of unintended irreversible actions.

Analysis suggests that specialization provides the solution instead of seeking a universal level of autonomy for a general-purpose agent. By deploying specialized agents with strictly scoped capabilities, organizations can achieve high reliability and safety for specific workflows. The following sections describe three common specialization patterns observed in production systems.

\subsubsection{Assistant Agents}
The most restrictive architectural pattern is the \textbf{Assistant Agent}, designed for summarization, extraction, and guidance. This agent type has read-only capabilities; it lacks interaction tools (such as \texttt{click} or \texttt{type}) or the ability to open new tabs. Assistant agents can read articles, extract data, or guide a user through complex platforms like AWS without the risk of modifying infrastructure. This pattern provides high utility with minimal risk and represents the majority of current safe production operations.

\subsubsection{Research Agents}
\textbf{Research Agents} require navigation and interaction capabilities to navigate through information. These agents typically utilize \texttt{navigate}, \texttt{click}, and \texttt{type} tools (often restricted to search bars). While necessary for gathering information, these capabilities introduce risks regarding prompt injection and unintended state changes.

Safety for research agents is enforced through domain allowlisting and programmatic constraints. For example, a research agent may be restricted to a specific set of domains (e.g., LinkedIn, public news sites) and prevented from executing actions on elements containing keywords like "send," "post," or "delete." This allows the agent to navigate and read content while preventing it from interacting with social features or messaging systems.

\subsubsection{Data Entry Agents}
\textbf{Data Entry Agents} are scoped to operate within a specific tab or workflow, often without the ability to navigate away from the target domain. These agents are permitted to enter data into specific forms but are restricted from other interactions. By limiting the agent's scope to a single context, the attack surface for prompt injection or hallucination is significantly reduced.

\subsection{Domain-Specific Configuration}

Beyond architectural specialization, production reliability improves through domain-specific configuration. This involves dynamically injecting operational rules or modifying the context based on the active domain.

One common pattern is dynamic prompt construction. When the agent enters a specific environment, such as Google Docs, the system appends a domain-specific instruction block to the system prompt. This block might detail best practices for that application, such as using keyboard shortcuts instead of UI buttons or using the \texttt{search\_replace} tool instead of manual editing.

A second pattern involves snapshot filtering for security. If an agent requires access to a platform like LinkedIn for recruiting but should not access personal messages, the execution layer can pre-process the accessibility tree snapshot to remove containers matching the "Messaging" or "inbox" selectors. This ensures that the sensitive data never reaches the LLM's context window, providing a guarantee that even a compromised LLM cannot leak private conversations.

\subsection{Summary}

Results confirm that browser agents can operate economically and reliably when architected correctly. The WebGames benchmark results show that current LLMs, supported by comprehensive tooling and context management, can achieve success rates approaching human performance on standard web tasks. However, the path to safe autonomous operation lies not in unrestricted general intelligence but in specialization, where architectural constraints and domain-specific configurations define the safety boundaries that LLMs alone cannot guarantee.

\section{Conclusion}
\label{sec:conclusion}

This paper demonstrates that production-ready browser agents require a fundamental architectural shift from general-purpose assistants to specialized tools. While current research often prioritizes universal web agents capable of performing any task, production experience reveals that reliability and safety result from architectural constraints instead of LLM scale.

Specialization provides a practical path for production operation that general-purpose systems struggle to replicate. Research agents do not require email capabilities; data entry agents do not require broad navigation privileges. By limiting agents to specific domains and actions through code as opposed to LLM judgment, reliability and security improve. This specialization enables autonomous operation by defining clear boundaries within which the agent can operate safely.

Browser agents perceive the web differently than humans. While humans process information sequentially through a viewport, requiring time to scroll and comprehend, agents process entire page structures instantly. This capability, when applied through hybrid accessibility and vision architectures, enables agents to operate with broader context awareness than human users. Future web standards may evolve to support this interaction mode with semantic annotations specifically designed for agents, similar to how ARIA attributes currently support assistive technologies.

However, security remains an evolving challenge. More intelligent and faster LLMs will certainly improve how browser agents operate, but prompt injection attacks will likely follow a cat-and-mouse pattern similar to phishing attacks on humans. Even highly intelligent humans fall victim to sophisticated phishing attempts; after training, they adapt and recognize those patterns until new, more creative attacks emerge. The same dynamic will apply to browser agents, new attack vectors will be discovered, mitigations will be developed, and the cycle continues. This ongoing security evolution reinforces why specialized agents with programmatic constraints are essential.

The critical question addresses how to architect systems that are useful yet safe. Internal evaluation on the WebGames benchmark demonstrates an approximately 85\% success rate across 53 challenges, compared to approximately 50\% reported for prior approaches. These results validate that architectural decisions, specifically regarding context management, execution layer design, and specialization, significantly impact reliability.

Browser agents offer productivity benefits when designed as purpose-built tools. The technology for autonomous operation exists today; the limiting factor is architecture, not LLM capability. The path forward lies in specialization.

\section*{Acknowledgments}

The production insights and benchmark results presented in this paper are derived from the author's work developing FillApp, a browser agent platform.

\vspace{2cm}

\bibliography{references}

@article{xie2024osworld,
  title={OSWorld: Benchmarking Multimodal Agents for Open-Ended Tasks in Real Computer Environments},
  author={Xie, Tianbao and Zhang, Danyi and Chen, Jixuan and Li, Xiaopeng and Zhao, Sida and Cao, Ruisheng and Gao, Tiffany and Pasupat, Panupong and Narasimhan, Karthik and Liang, Percy and others},
  journal={arXiv preprint arXiv:2404.07972},
  year={2024}
}

@article{koh2024visualwebarena,
  title={VisualWebArena: Evaluating Multimodal Agents on Realistic Visual Web Tasks},
  author={Koh, Jing Yu and Lo, Robert and Jang, Lawrence and Duvvur, Vikram and Lim, Ming Chong and Huang, Po-Yu and Neubig, Graham and Zhou, Shuyan and Salakhutdinov, Ruslan and Fried, Daniel},
  journal={arXiv preprint arXiv:2401.13649},
  year={2024}
}

@article{webgames2025,
  title={WebGames: Challenging General-Purpose Web-Browsing AI Agents},
  author={Thomas, George and Chan, Alex J. and Kang, Jikun and Wu, Wenqi and Christianos, Filippos and Greenlee, Fraser and Toulis, Andy and Purtorab, Marvin},
  journal={arXiv preprint arXiv:2502.18356},
  year={2025}
}

@article{shapira2025mindtheweb,
  title={Mind the Web: The Security of Web Use Agents},
  author={Shapira, Avishag and Weiss, Gal and Bitton, Ron and Inbar, Eliya and Nassi, Ben and Elovici, Yuval},
  journal={arXiv preprint arXiv:2506.07153},
  year={2025}
}

@article{cohen2025fuzzing,
  title={In-Browser LLM-Guided Fuzzing for Real-Time Prompt Injection Testing in Agentic AI Browsers},
  author={Cohen, Avihay},
  journal={arXiv preprint arXiv:2510.13543},
  year={2025}
}

@online{openai2025atlas,
  author = {{OpenAI}},
  title = {ChatGPT Atlas: AI-First Browser with Agent Mode},
  year = {2025},
  month = {October},
  url = {https://openai.com/blog/chatgpt-atlas}
}

@online{perplexity2025comet,
  author = {{Perplexity AI}},
  title = {Introducing Comet: AI-Powered Browser for Research and Automation},
  year = {2025},
  month = {July},
  url = {https://www.perplexity.ai/blog/comet-browser}
}

@online{anthropic2024computeruse,
  author = {{Anthropic}},
  title = {Introducing Computer Use: A New Capability for Claude},
  year = {2024},
  month = {October},
  url = {https://www.anthropic.com/news/computer-use}
}

@online{anthropic2024chrome,
  author = {{Anthropic}},
  title = {Claude for Chrome: Browser Automation Extension},
  year = {2024},
  url = {https://chrome.google.com/webstore/detail/claude-for-chrome}
}

@article{he2024webvoyager,
  title = {WebVoyager: Building an End-to-End Web Agent with Large Multimodal Models},
  author = {He, Hongliang and Yao, Wenlin and Ma, Kaixin and Yu, Wenhao and Dai, Yong and Zhang, Hongming and Lan, Zhenzhong and Yu, Dong},
  journal = {arXiv preprint arXiv:2401.13919},
  year = {2024}
}

@misc{playwrightmcp2024,
  author = {{Microsoft}},
  title = {Playwright MCP: Model Context Protocol Server for Browser Automation},
  year = {2024},
  howpublished = {\url{https://github.com/microsoft/playwright-mcp}}
}

@misc{chromedevtoolsmcp2024,
  author = {{Model Context Protocol Team}},
  title = {Chrome DevTools MCP: Model Context Protocol for Chrome Browser Control},
  year = {2024},
  howpublished = {\url{https://github.com/modelcontextprotocol/servers/tree/main/src/chrome-devtools}}
}

@misc{browseruse2024,
  author = {{Browser Use Team}},
  title = {Browser Use: Make Websites Accessible for AI Agents},
  year = {2024},
  howpublished = {\url{https://github.com/browser-use/browser-use}}
}

@techreport{w3c2018wcag,
  author = {{W3C}},
  title = {Web Content Accessibility Guidelines (WCAG) 2.1},
  institution = {World Wide Web Consortium},
  year = {2018},
  type = {W3C Recommendation},
  url = {https://www.w3.org/TR/WCAG21/}
}

@online{webaim2024million,
  author = {{WebAIM}},
  title = {The WebAIM Million: An Annual Accessibility Analysis of the Top 1,000,000 Home Pages},
  year = {2024},
  url = {https://webaim.org/projects/million/}
}

@techreport{chrome2024devtools,
  author = {{Google Chrome Team}},
  title = {Chrome DevTools Protocol: Accessibility Domain},
  institution = {Google},
  year = {2024},
  url = {https://chromedevtools.github.io/devtools-protocol/tot/Accessibility/}
}

@online{brave2025comet,
  author = {Sampayo, Javier and {Brave Security Team}},
  title = {Prompt Injection Attacks on Perplexity's AI Browser Comet},
  year = {2025},
  url = {https://brave.com/blog/comet-prompt-injection/}
}

@article{liu2024prompt,
  title = {Prompt Injection Attacks and Defenses in LLM-Integrated Applications: A Comprehensive Review},
  author = {Liu, Yupei and Deng, Yihao and Song, Dawn},
  journal = {arXiv preprint arXiv:2310.12815},
  year = {2024}
}

@online{openai2024caching,
  author = {{OpenAI}},
  title = {Prompt Caching},
  year = {2024},
  url = {https://platform.openai.com/docs/guides/prompt-caching}
}

@online{google2025geminivision,
  author = {{Google DeepMind}},
  title = {Gemini API: Vision},
  year = {2025},
  url = {https://ai.google.dev/gemini-api/docs/vision}
}

\end{document}